\documentclass[aps,pre,preprint,numerical,superscriptaddress,showkeys,showpacs,floatfix]{revtex4-1}
\usepackage{graphicx}
\usepackage{subfig}
\usepackage{amsmath,mathrsfs,amssymb}
\begin{document}
\title{Self-similar Turing Patterns: An Anomalous diffusion consequence}
\author{D. Hern\'andez}
\affiliation{Maestr\'ia en Ciencias de la Complejidad, Universidad Aut\'onoma de la Ciudad de M\'exico, Ciudad de M\'exico, Laboratorio Nacional de Ciencias de la Complejidad, Ciudad de M\'exico}
\author{E. C. Herrera-Hern\'andez}
\email{erik.herrera@cidesi.edu.mx}
\affiliation{CONACYT-Centro de Ingenier\'ia y Desarrollo Industrial, Av. Playa pie de la cuesta 702, Desarrollo Sn. Pablo, 76125, Quer\'etaro, Qro., Mexico}
\author{ M.N\'u\~nez-L\'opez}
\affiliation{Departamento de Matem\'aticas Aplicadas y Sistemas, DMAS
Universidad Aut\'onoma Metropolitana, Cuajimalpa}
\author{H.Hern\'andez Coronado}
\affiliation{Departamento de F\'isica, Facultad de Ciencias, UNAM,  A. P. 50-542, Mexico DF, 04510, Mexico.}
\begin{abstract}
In this work we show that under specific anomalous diffusion conditions, chemical systems can produce well-ordered self-similar concentration patterns through a diffusion-driven instability. We also find spiral patterns and patterns with  mixtures of rotational symmetries not reported before.\\
The type of anomalous diffusion discussed in this work, either subdiffusion or superdiffusion, is a consequence of the medium heterogeneity, and is modelled through a space-dependent diffusion coefficient with a power-law functional form.
\end{abstract}
\maketitle
\section{Introduction}\label{sec:1}
Over the last hundred years, a vast amount of experimental evidence about anomalous transport processes has accumulated in the scientific literature ~\cite{Bouchaud, Restaurant, R.Metzler}. This triggered a lot of theoretical research aiming to understand and predict such processes, and as a consequence, a great variety of mathematical models for anomalous diffusion have been proposed. Among these models, the most successful in terms of predictions and  wide range of applications are the continuous time random walk formalism, the L\'evy flights model, and the fractional generalizations of the advection-diffusion equation ~\cite{Metzler}. Such models are specially well suited for describing sub and superdiffusion due to scale free waiting times and jump lengths in the diffusing particle dynamics, but give poor results for anomalous diffusion due to long range correlations in the particle displacements ~\cite{DamUACM,Matheron}. This type of anomalous diffusion (superdiffusion or subdiffusion caused by long-range correlation) is ubiquitous in pre-fractal and heterogeneous porous media ~\cite{Havlin,Sahimi}, and a considerable fraction of the models found in the literature were developed to describe it. To our knowledge,  the first generalization of the diffusion equation for  fractal geometries was proposed by O'Shaughnessy and Procaccia ~\cite{Procaccia}, and although they did not solved the problem, they gave a good approximation for the  probability distribution function for finding a diffusing particle at a distance $r$ from the origin at time $t$, for specific fractal domains. This led the scientific community to develop several generalizations of the diffusion equation for describing diffusion in pre-fractal geometries ~\cite{Campos,Metzler-Glockle}, and even though this is still an open problem, some of these models are valid models for anomalous diffusion and a good starting point to study the consequences that this transport process has in phenomena where diffusion plays an important role. As an example of this, think of diffusion limited chemical reactions and all the interesting dynamical states exhibited by these systems: how does anomalous diffusion affect their dynamic behavior or the conditions for their appearance? Is it possible for chemical systems to generate exotic emergent structures under anomalous diffusion conditions? In this context, there are some studies about the interplay between anomalous diffusion and Turing patterns, and also between anomalous diffusion and wave-front propagation in chemical systems ~\cite{Mendez,Hernandezwave}; however in the majority of these cases anomalous diffusion was modelled either with the continuous time random walk formalism or with fractional derivatives ~\cite{Henry,Barrios,DHernandez,Mendez}. Although these approaches are interesting from a mathematical point of view and because their potential physicochemical implications, like weaker conditions for  diffusion driven instabilities in a system of two chemicals under subdiffusion, or Turing patterns moving with constant velocity, they  still need experimental verification, and some of them, a derivation form first principles.\newline
On the other hand, the essential ingredient in many of the generalized diffusion equations for transport in pre-fractal geometries or highly heterogeneous porous media, and in particular the model put forward by O'Shaughnessy and Procaccia, is a space-dependent diffusion coefficient with a power-law functional form ~\cite{Procaccia,Hector2012,Erik2013,Hernandez,Chang}. In this work, we explore the effects of this type of diffusion coefficients (and therefore, of this type of anomalous diffusion) in the context of pattern formation in reaction-diffusion systems; more precisely, in the emergence of concentration patterns generated through a diffusion driven instability.\newline
The remainder of the paper is organized as follows. As a part of the introduction, Subsection \ref{ssec:1}, we deduce the generalized diffusion equation we  use along the paper and discuss some of its properties. In Section \ref{sec:2} we present the reaction-diffusion model and find  the necessary and sufficient conditions for a diffusion driven instability along with the  generalized dispersion relation. In Section \ref{sec:3} we present and discuss the numerical results for normal diffusion, subdiffusion and superdiffusion. Finally, in Section \ref{sec:4} we draw some conclusions about this work.\\
\subsection{Generalized Diffusion Equation} \label{ssec:1}
In the absence of chemical reactions, the concentration of any chemical species must fulfill the local conservation law given by the continuity equation
\begin{equation}\label{eq:Econt}
\frac{\partial u}{\partial t}=-\nabla \cdot J,
\end{equation}
where $J(x,t)=-D(x)\nabla u(x,t)$ denotes the particle flux at point $x \in \mathbb{R}^n $ at time $t$ and $n \in \mathbb{N}^n$;  $D(x)$ represents a space-dependent diffusion coefficient and $u(x,t)$ the local concentration of a chemical species at time $t$.\\
We consider the spatial region where diffusion takes place as a two dimensional disk, and  assume that the medium heterogeneities are only along the radial coordinate. Therefore, the diffusion coefficient must be a function of the radial coordinate only, and by hypothesis is given by 
\begin{equation}\label{eq:Cdif}
D(r)=D_{u}r^{-\lambda},
\end{equation}
with $D_{u}$ a constant with the appropriate physical units.\\
Writing the particle flux in polar coordinates 
\begin{equation}\label{eq:Flujo}
J(r,\theta)=-D_{u} r^{-\lambda}\Bigl( \frac{\partial u}{\partial r}\hat{e}_r+\frac{1}{r}\frac{\partial u}{\partial \theta}\hat{e}_{\theta}\Bigr),
\end{equation}
(with $\hat{e}_r $ and $\hat{e}_{\theta}$  unitary vectors along the radial and azimuthal directions respectively), and substituting it in Eq.(\ref{eq:Econt}), we arrive at the following two dimensional generalized  diffusion equation
\begin{eqnarray} \label{eq:Ediff}
\frac{\partial u}{\partial t}=D_{u}\nabla^2_{\lambda}u\\
\nabla^2_{\lambda}u=\frac{1}{r}\frac{\partial}{\partial r}\Bigl( r^{1-\lambda}\frac{\partial u}{\partial r} \Bigr)+\frac{1}{r^{2+\lambda}}\frac{\partial^2 u}{\partial \theta^2}. \nonumber
\end{eqnarray}
The first important thing to notice about Eq.(\ref{eq:Ediff}) is that it implies an anomalous behaviour for the mean square displacement for the chemical species $u$; this is shown in the following expression ~\cite{Procaccia}
\begin{equation}\label{eq:MSD}
\langle r(t)^2\rangle \propto \int_0^{2\pi}\int_0^{\infty}r^2u(r,\theta,t)rdrd\theta \propto t^{\frac{2}{2+\lambda}}.
\end{equation}
From  Eq.(\ref{eq:MSD}) it follows that for positive values of $\lambda$, Eq.(\ref{eq:Ediff}) describes a subdiffusive regime, whereas for negative values of such parameter a superdiffusive regime is implied. At this point, it is important to mention that, when equations of the type Eq.(\ref{eq:Ediff}) has been proposed in the literature for studying diffusion in domains with pre-fractal structure, the parameter $\lambda$ is always taken positive ~\cite{Chang}, and when deduced in such a context, it also takes positive values ~\cite{Havlin,Hernandez}. In our case we  consider that $\lambda$  takes either positive or negative values, implying that in this work, Eq.(\ref{eq:Ediff}) is not intended to describe diffusion in domains with fractal structure, instead it is taken as a model of anomalous diffusion for heterogeneous domains.\\
When Eq.(\ref{eq:Ediff}) is posed as an initial boundary value problem with homogeneous boundary conditions, the eigenfunctions associated to each of the differential operators can be obtained using separation of variables, {\it i.e.} by assuming a solution of the form $u(r,\theta,t)=T(t)f(r)g(\theta),\quad \theta \in [0,2\pi)$, it follows that the radial, temporal and angular functions satisfy the eigenvalue equations
\begin{eqnarray}\label{eq:EigEq.}
\frac{1}{D_{u}}\frac{dT}{dt}=c_1T, \quad \frac{d^2g}{d\theta^2}=-c_2g,\\ 
r^2\frac{d^2f}{dr^2}+\omega r\frac{df}{dr}+r^{2+\lambda}c_1f=c_2f,
\end{eqnarray}
where $\omega=1-\lambda$ and $c_i$ with $i=1,2$, are constants of separation given by the following expressions: $c_1=-k^\nu$, $c_2=m^2,m\in \mathbb{Z}$. The constants $k$ and $\nu$ are determined by the boundary conditions and by writing the radial differential equation as a Bessel equation through the  following set of transformations: $ r=x^{\kappa}$, $f=x^{\zeta}\bar{f}$ and $y=\kappa k^{\nu/2}x$, where $\kappa= 2/(2+\lambda)$, $\zeta=(1-\omega)/(2+\lambda)$ and $\nu=2+\lambda$. After such changes of variables, the radial differential equation takes the form
\begin{equation}\label{eq:Bessel}
\frac{\partial^2 \bar{f}}{\partial y^2}+y^{-1}\frac{\partial \bar{f}}{\partial y}+ \Bigl(1- \Omega^2 y^{-2} \Bigr)\bar{f}=0,
\end{equation}
where $\Omega^2=\frac{4m^2+\lambda^2}{(2+\lambda)^2} $. Notice that $\Omega \in \mathbb{Z}$ only when $\lambda=0$ (normal diffusion), meaning that the general solution of Eq.(\ref{eq:Bessel}) in terms of the radial coordinate is given by
\begin{eqnarray}\label{eq:SolBess}
\bar{f}(r)&=&A_1J_{\Omega}\Bigl( \rho(kr)  \Bigr)+ A_2J_{-\Omega}\Bigl( \rho(kr)  \Bigr), \\
\rho(kr)&=&\frac{2}{2+\lambda}\left[kr\right]^{(2+\lambda)/2}, \nonumber
\end{eqnarray}
where $A_1$  and $A_2$ are constants and the symbols $J_{\Omega}(\xi)$ and $J_{-\Omega}(\xi)$ denote  Bessel functions of order $\Omega$ of the first and second kind respectively. Because we are interested in bounded solutions $A_2=0$;  the solutions of the  radial eigenvalue problem are
\begin{eqnarray}\label{eq:SolBess2}
f(r)&=&A_1r^{\lambda/2}J_{\Omega}\Biggl(\frac{2}{2+\lambda}[kr]^{\frac{2+\lambda}{2}}  \Biggr)\\
\Omega &=&\sqrt{\frac{4m^2+\lambda^2}{(2+\lambda)^2}}. \nonumber
\end{eqnarray}
The possible values of $k$ in Eq.(\ref{eq:SolBess2}) are determined through the homogeneous boundary value conditions. For a domain given by an unitary circle with Neumann homogeneous boundary conditions: $k=k_{m,n}=\Bigl(\frac{2+\lambda}{2} j_{m,n}\Bigr)^{2/[2+\lambda]}$, where $j_{m,n}$ is the $n$-th root of the following expression
\begin{equation}\label{eq:root}
\left(2+\lambda\right)\frac{dJ_{\Omega}(x)}{dx}x+\lambda J_{\Omega}(x)=0.
\end{equation}
Notice that for normal diffusion ($\lambda=0$), Eq.(\ref{eq:root}) reduces to the appropriate zero flux boundary conditions with $x=k$. Henceforth, the eigenfunctions of Eq.(\ref{eq:Ediff}) in an unitary circle with Neumann homogeneous boundary conditions are
\begin{equation}\label{eq:eigdif}
\phi_{m,n}(r,\theta,t)\propto e^{-D_{u}k_{m,n}^{\left(2+\lambda\right)}t}r^{\lambda/2}J_{\Omega}\Biggl(\frac{2}{2+\lambda}[k_{m,n}r]^{\frac{2+\lambda}{2}}  \Biggr)e^{im\theta}+c.c. \quad .
\end{equation}
\section{The Model}\label{sec:2}
Consider a reaction-diffusion system of the form
\begin{eqnarray}\label{eq:RDG}
\frac{\partial u}{\partial t}&=&D_u\nabla^2_{\lambda}u+F(u,v) \\
\frac{\partial v}{\partial t}&=&D_v\nabla^2_{\lambda}v+G(u,v), \nonumber
\end{eqnarray}
defined in a circular domain of radius $r_b$ with zero flux boundary conditions. Taking the reaction terms as those corresponding to the {\bf BVAM} model ~\cite{BarrioNum} and writing Eq.(\ref{eq:RDG}) in  dimensionless form we have the following expressions
\begin{eqnarray}\label{eq:RD}
\frac{\partial u}{\partial t}&=&\delta D\nabla^2_{\lambda}u+\alpha u(1-r_1v^2)+v(1-r_2u)\\ 
\frac{\partial v}{\partial t}&=&\delta \nabla^2_{\lambda}v+\beta v(1+\left[\alpha r_1/\beta\right]uv)+u(r_2v-\alpha), \nonumber
\end{eqnarray}
\begin{equation}
r\in [0,1]; \quad \frac{\partial u(r,\theta,t)}{\partial r}\biggl|_{r=1}=0, \quad \frac{\partial v(r,\theta,t)}{\partial r}\biggl|_{r=1}=0,
\end{equation}
where $D=D_u/D_v$ and  $\delta=\delta_0 r_b^{-(2+\lambda)}$ with $\delta_0$ a positive constant.\\
Although without a single mention of anomalous diffusion, a particular case of Eq.(\ref{eq:RDG})  was proposed by Berding {\it et al.} ~\cite{Berding} to explain the pre-pattern formation mechanism for the spiral type patterns of the Sunflower head. In their model $\lambda=-2$ and the kinetics is taken to be the well known Gierer-Meinhardt chemical reaction kinetics ~\cite{Gierer}. In this respect, it is interesting to mention   their argument to justify the spatial dependence of the diffusion coefficients:  ``{...\it it could be motivated by, for example, some inhomogeneous cell densities, which make the diffusion coefficients increasing functions of distance from the centre}". In our case, instead of fixing the value of $\lambda$, we explore the effects it has over Turing patterns through Eq.(\ref{eq:RD}). In the following sections we show the results of such exploration.\\
\subsection{Model Linear Analysis}
\begin{figure}[ht]
\subfloat{\includegraphics[scale=0.7]{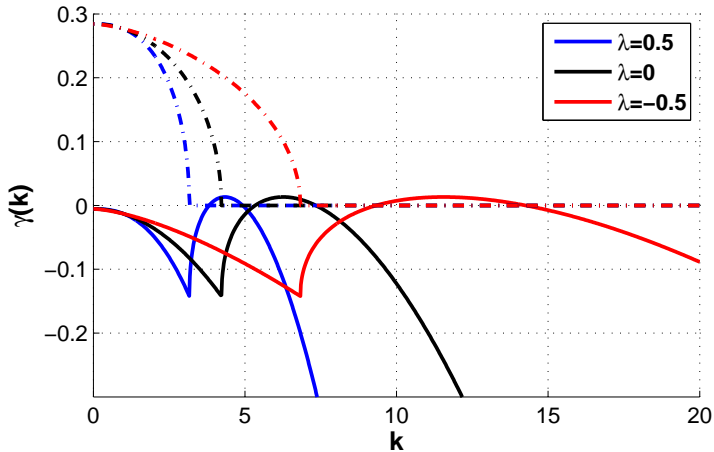}}
\caption{ (Color online) Real (continuous lines) and imaginary part (dotted lines) of dispersion relation for different values of $\lambda$. The parameter values used for this figure are $\alpha =0.899$, $\beta =-0.91$, $\delta =0.01011$, $D =0.516$.}
\label{fig:1}
\end{figure}

Equation (\ref{eq:RD}) has a unique stationary uniform solution given by $(u_0,v_0)=(0,0)$. The linearized equations around such state are
\begin{eqnarray}\label{eq:RDl}
\frac{\partial u}{\partial t}&=&\delta D\nabla^2_{\lambda}u+\alpha u+v\\
\frac{\partial v}{\partial t}&=&\delta \nabla^2_{\lambda}v+\beta v-\alpha u. \nonumber
\end{eqnarray}
Using the results discussed in subsection (\ref{ssec:1}), it is easy to see that the temporal evolution of the eigenmodes associated to Eq.(\ref{eq:RDl}) is given by:  $u=U_0\exp(\gamma t)f(r)e^{im\theta}$ and $v=V_0\exp(\gamma t)f(r)e^{im\theta}$, where the radial function is shown in Eq.(\ref{eq:SolBess2}) and  $\gamma$ is a solution of the following eigenvalue equation
\begin{equation}\label{eq:reld}
\gamma
\begin{pmatrix}
U_0 \\
V_0
\end{pmatrix}=
\begin{pmatrix}
-D\delta k^{2+\lambda}+\alpha & 1\\
-\alpha & -\delta k^{2+\lambda}+\beta
\end{pmatrix}
\begin{pmatrix}
U_0\\
V_0
\end{pmatrix}.
\end{equation}
After solving Eq.(\ref{eq:reld}) for  $\gamma$, from now on called dispersion relation, we arrive at the following expression
\begin{eqnarray}\label{eq:gamma}
\gamma(k)&=&\Gamma_1(k)+\Gamma_2(k); \nonumber \\
\Gamma_1(k)&=&\frac{\alpha+\beta-\delta k^{2+\lambda}(D+1)}{2},  \nonumber\\
\Gamma_2(k)&=&\frac{\sqrt{[2\Gamma_1(k)]^2-4h(k)}}{2}, \\
h(k)&=&\delta^2D k^{(4+2\lambda)}-(\beta D+\alpha)\delta k^{2+\lambda}+\alpha(\beta+1). \nonumber
\end{eqnarray}
Through Eq.(\ref{eq:gamma}), Turing conditions for a diffusion-driven instability can be written as: $\Gamma_1(0)<0$ and $h(k)<0$ for $k\in (k_1,k_2)$, with $k_1$ y $k_2$ solutions of  $h(k)=0$. This numbers define the instability window and their explicit forms are
\begin{eqnarray}\label{eq:Zerosh}
k_1&=&\Biggl( \frac{\beta D+\alpha}{2\delta D}-\frac{\sqrt{(\beta D+\alpha)^2-4 D\alpha(\beta+1) }}{2\delta D}\Biggr)^{1/(2+\lambda)} \nonumber \\
k_2&=&\Biggl( \frac{\beta D+\alpha}{2\delta D}+\frac{\sqrt{(\beta D+\alpha)^2-4 D\alpha(\beta+1) }}{2\delta D}\Biggr)^{1/(2+\lambda)}.
\end{eqnarray}
Taking $u$ as the activator and $v$ as the inhibitor, Turing conditions written in terms of the model parameters take the following form
\begin{equation}\label{eq:Tc}
\begin{split}
 &\alpha+\beta <0, \quad \alpha (\beta+1)>0, \quad \beta D+\alpha >0, \quad  D  < 1.
\end{split}
\end{equation}
The critical numbers $k_c$ and $D_c$, corresponding to the value of $k$ at which the real part of the dispersion relation has a maximum and the point at which the instability first appears (bifurcation point), are 
\begin{equation} \label{eq:Critv}
k_c=\biggl(\frac{\beta D+\alpha}{2\delta D}\biggr)^{1/(2+\lambda)},
D_c=\frac{\alpha}{\beta^2}\left(2+\beta -2\sqrt{\beta+1}\right).
\end{equation}
Assuming the initial condition for Eq.(\ref{eq:RD}) as a small perturbation around the stationary steady sate, allows to write the general solution of  Eq.
(\ref{eq:RDl}) as
\begin{eqnarray}\label{eq:Sol}
u(r,\theta,t)=\sum_{m}\sum_{n}A_{m,n}e^{\gamma(k_{m,n})t}r^{\lambda/2}J_{\Omega(m)}\Bigl(\rho(k_{m,n}r)\Bigr)e^{im\theta}+c.c. \nonumber\\
v(r,\theta,t)=\sum_{m}\sum_{n}B_{m,n}e^{\gamma(k_{m,n})t}r^{\lambda/2}J_{\Omega(m)}\Bigl(\rho(k_{m,n}r)\Bigr)e^{im\theta}+c.c.,
\end{eqnarray}
where the constants $A_{m,n}$ and $B_{m,n}$ are defined in terms of the initial conditions, and $k_{m,n}=\Bigl(\frac{2+\lambda}{2} j_{m,n}\Bigr)^{2/[2+\lambda]}$ with $j_{m,n}$  the $n$-th root of Eq.(\ref{eq:root}).\\
Notice that for each number $k_{m,n}$ lying inside the instability window, {\it i.e.} whenever they lie between the numbers shown in Eq.(\ref{eq:Zerosh}), the corresponding eigenmode will grow exponentially fast until the nonlinear terms comes into play and saturate its growth.\\
On the other hand, notice that Eq.(\ref{eq:Zerosh}) implies that for fixed chemical kinetic parameters, the instability window has different sizes for subdiffusion $(\lambda > 0)$, normal diffusion $(\lambda = 0)$ and superdiffusion $(\lambda < 0)$, taking its smallest size in the subdiffusive case and its largest size in the superdiffusive regime. This result means that for fixed chemical kinetics parameters, it is easier to have several unstable eigenmodes in the superdiffusive regime than in the normal and subdiffusive cases, implying that Turing patterns with mixed symmetries are more likely to occur under superdiffusion conditions. An example of the dispersion relation behavior for different diffusive regimes is shown in Fig.(\ref{fig:1})
\section{Numerical Results}\label{sec:3}
Taking a previous study of Turing patterns in the unitary disc under normal diffusion conditions as guideline ~\cite{Barrio}, in this section we study how anomalous diffusion affects Turing patterns and their symmetries as a function of disc size.\\
The numerical results shown in this section were obtained through an explicit finite difference scheme based on Euler integration method, with zero-flux boundary conditions; the initial conditions were taken as  small random perturbations around the system stationary steady state. For an explicit discussion about the numerical integration method see ~\cite{Barrio}. \\
The  results of this section were obtained with the following values for the model parameters: $\alpha=0.899$, $\beta=-0.91$, $D=0.516$, $r_1=r_2=0.2$. The values for the remaining parameters are discussed in the text.
\subsection{Normal diffusion and subdiffusion}\label{ssec:2}
\begin{figure}[ht]
\centering
\subfloat{\includegraphics[scale=0.5]{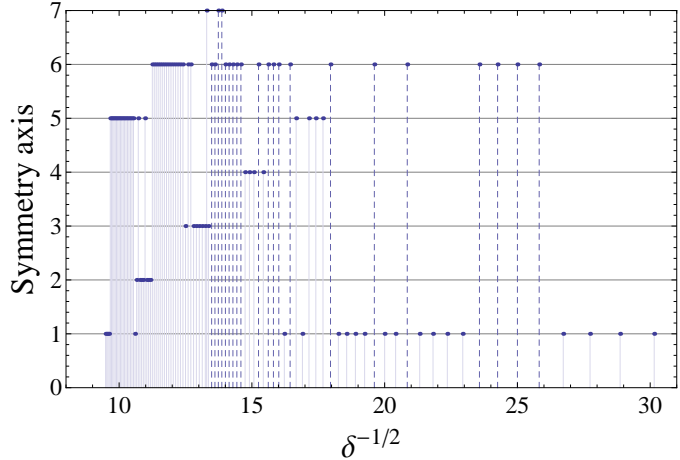}}
\caption{(Color online) Turing pattern rotational symmetries for normal diffusion. The dashed lines denote noncentrosymmetric patterns, while the continuous one represents centrosymmetric patterns.}
\label{fig:2}
\end{figure}

\begin{figure} 
\centering
\subfloat[$\delta=0.0036$] {\includegraphics[scale=0.7]{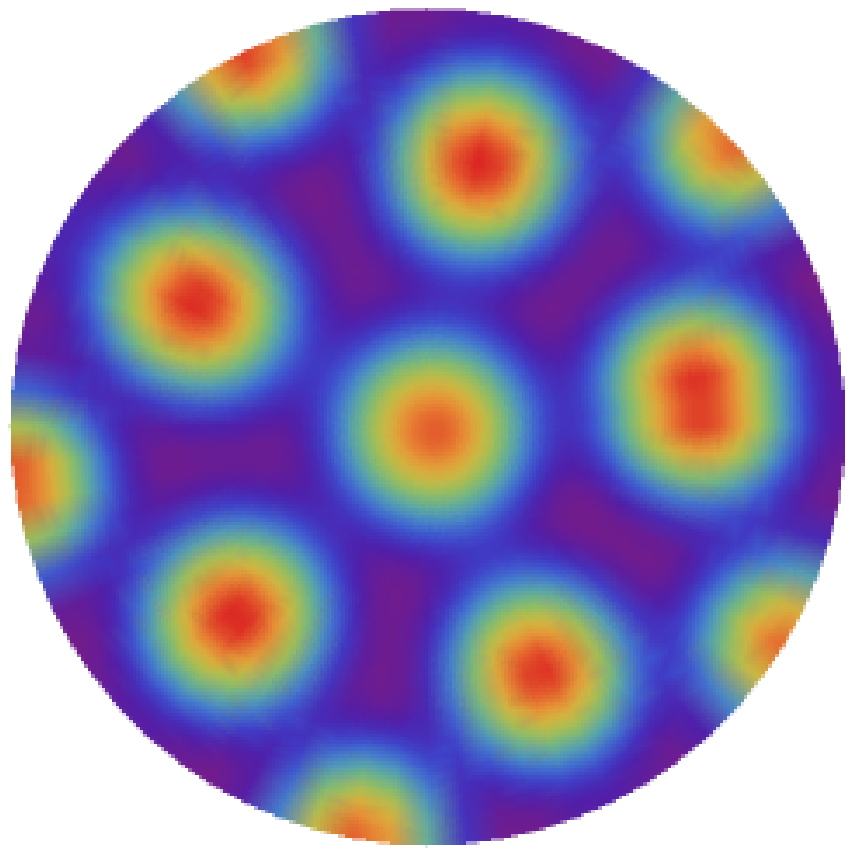}}\hspace{0.6cm}
\subfloat[$\delta=0.0070$] {\includegraphics[scale=0.7]{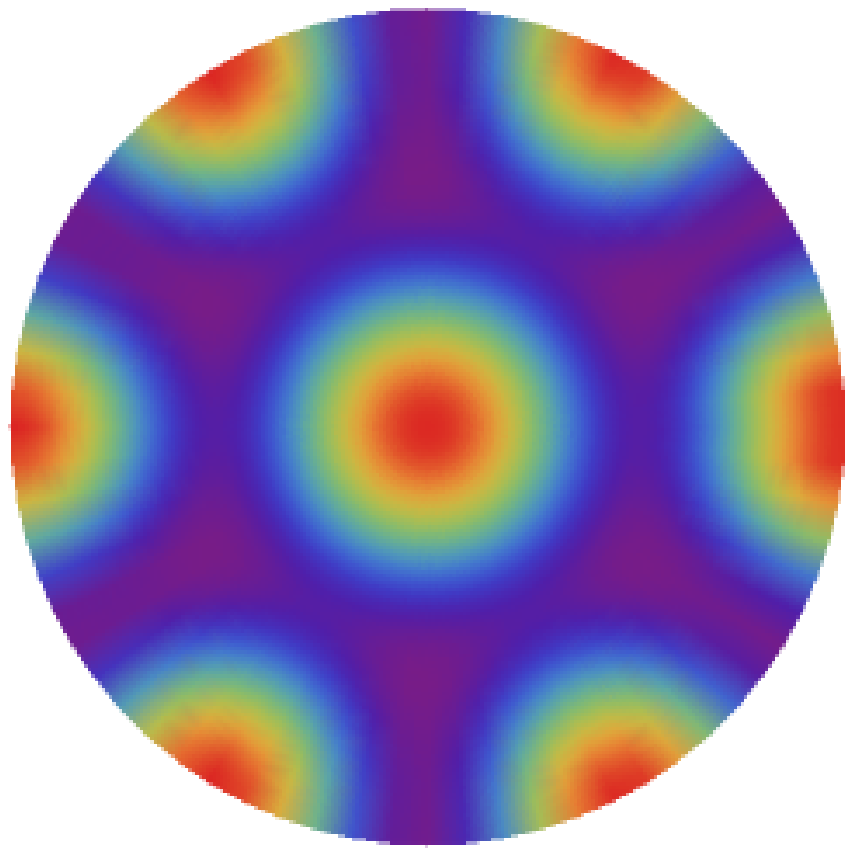}}\\
\subfloat[$\delta=0.0086$] {\includegraphics[scale=0.7]{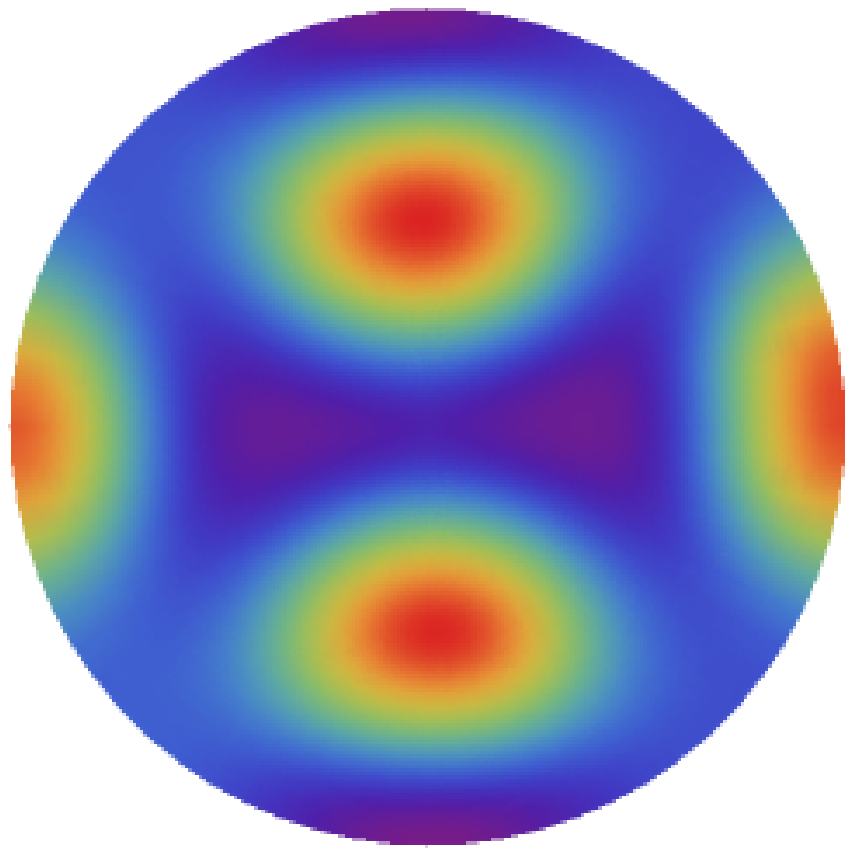}}\hspace{0.6cm}
\subfloat[$\delta=0.0044$] {\includegraphics[scale=0.7]{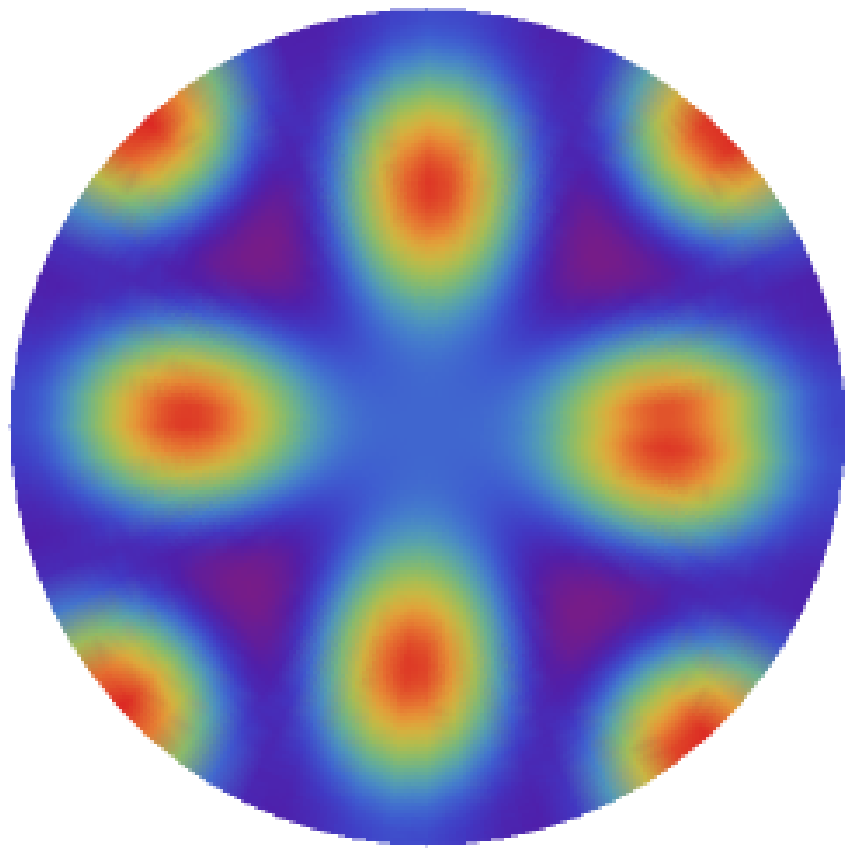}}
\caption{(Color online) Centrosymmetric  Turing patterns  under normal diffusion conditions, $ \lambda = 0$. {\bf a)} Pentagonal Turing pattern.  {\bf b)} Hexagonal Turing pattern.   {\bf c)}  Two-fold symmetric Turing pattern. {\bf d)} Tetra-symmetric Turing pattern.}
\label{fig:3}
\end{figure}

\begin{figure} 
\centering
\subfloat {\includegraphics[scale=0.4]{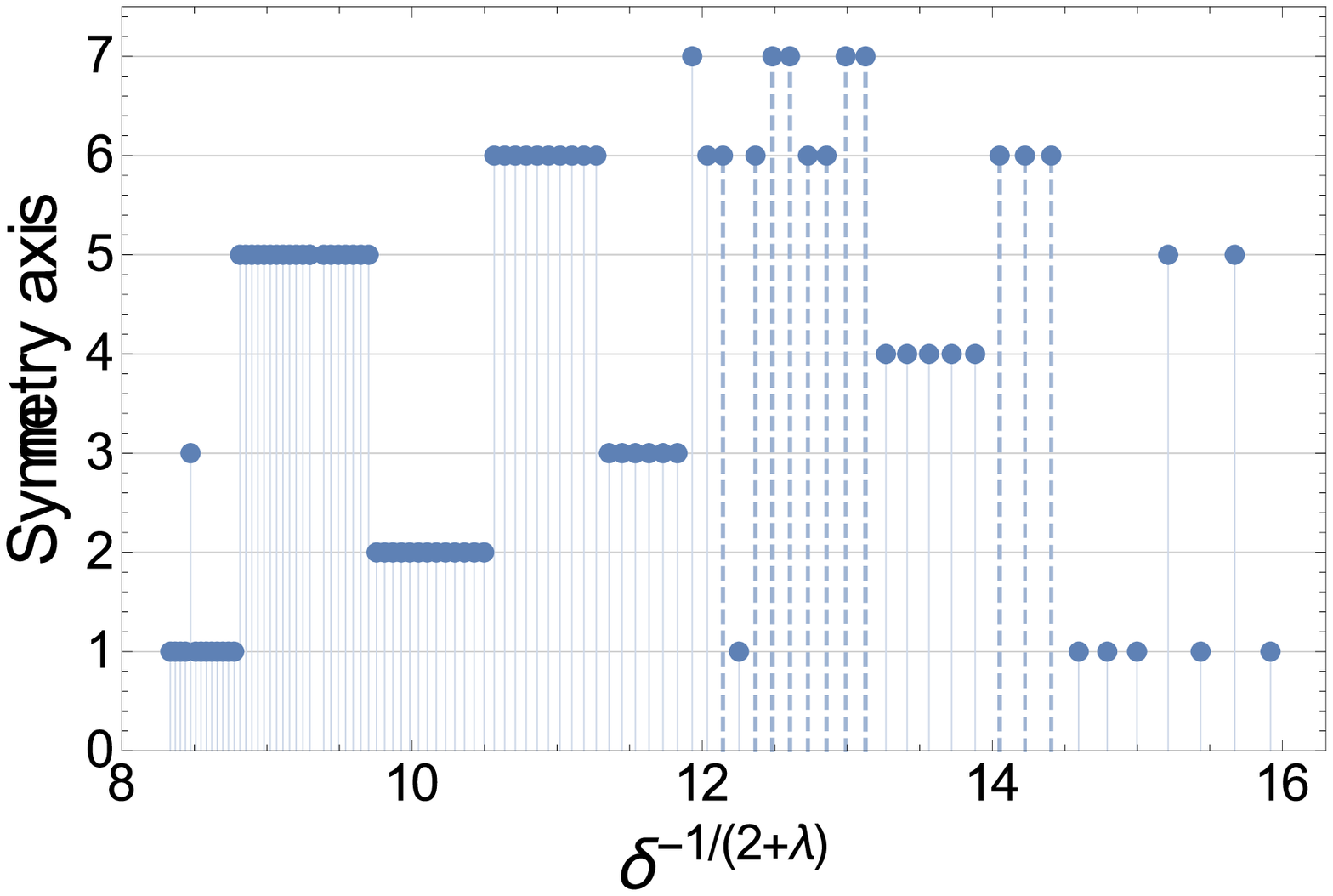}}\hspace{0.6cm}
\subfloat {\includegraphics[scale=0.4]{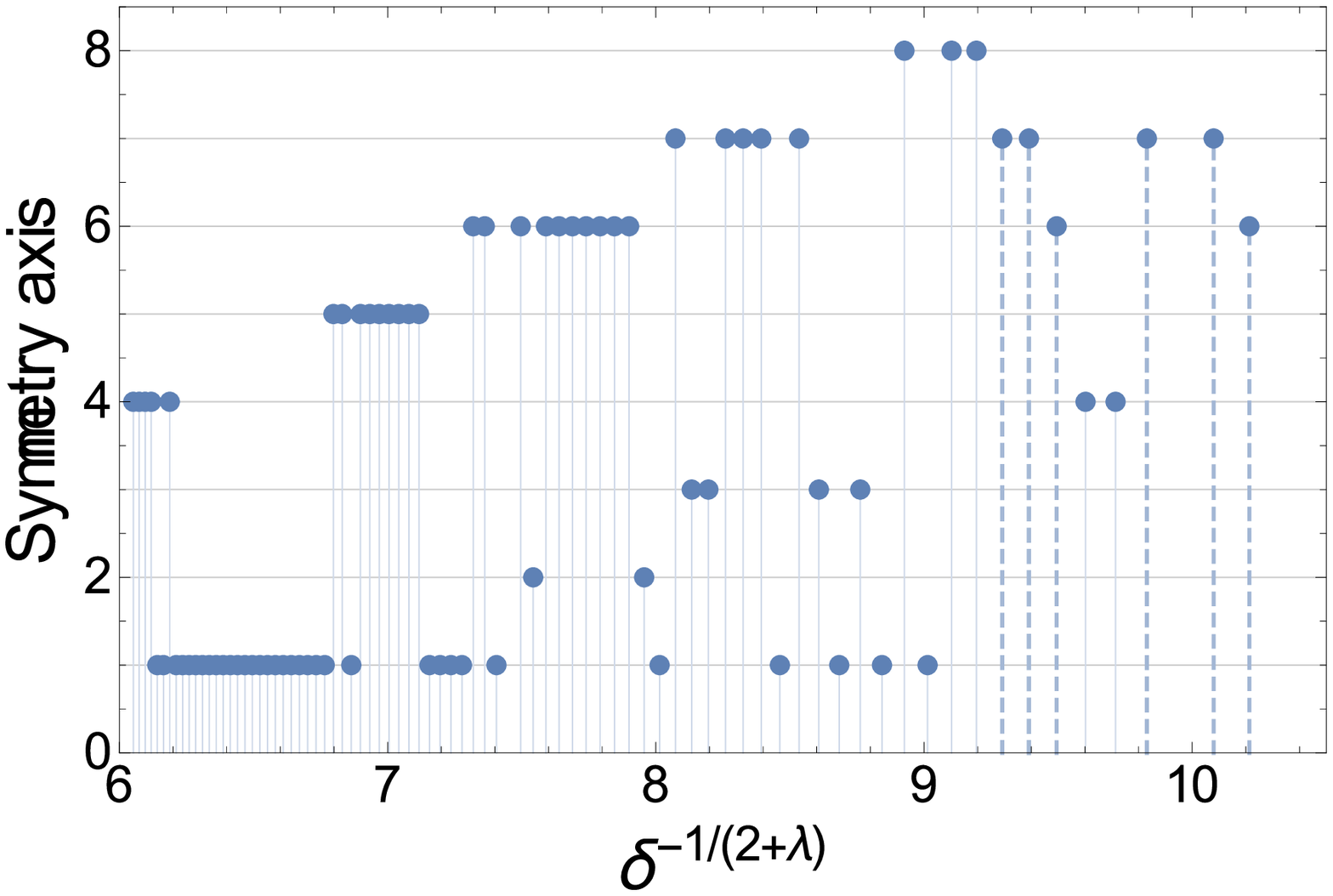}}
\caption{Turing pattern rotational symmetries for subdiffusive regimes. {\bf a)} Rotational symmetries for $\lambda=0.1$. {\bf b)} Rotational symmetries for $\lambda=0.5$} 
\label{fig:4}
\end{figure}

\begin{figure} 
\centering
\subfloat[$\delta=0.0053$] {\includegraphics[scale=0.7]{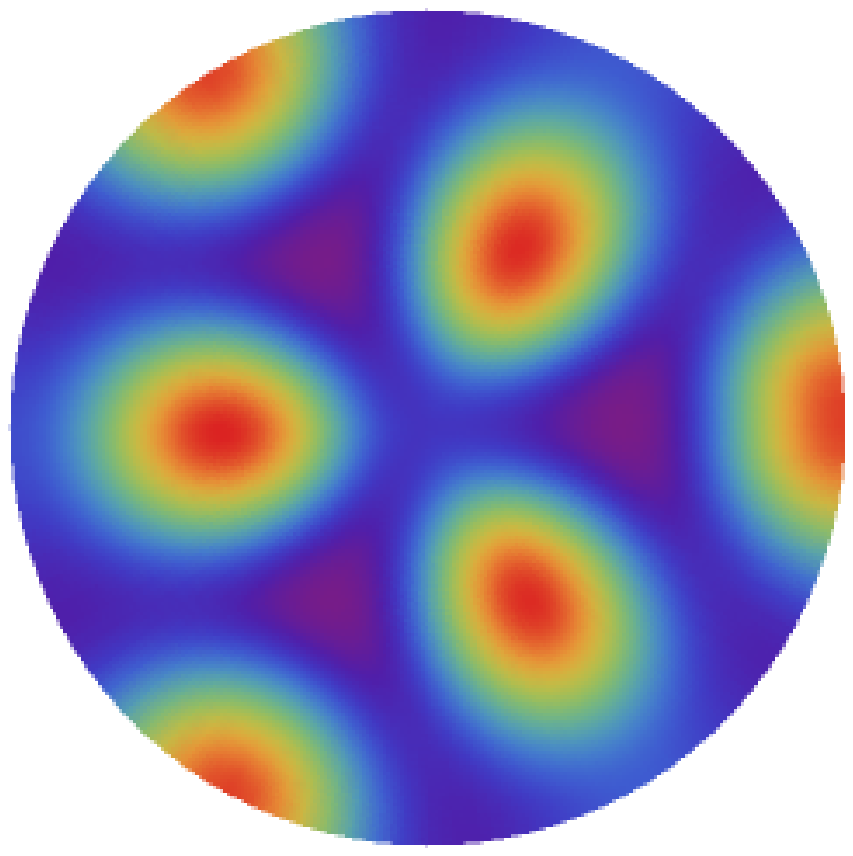}}\hspace{0.6cm}
\subfloat[$\delta=0.008$] {\includegraphics[scale=0.7]{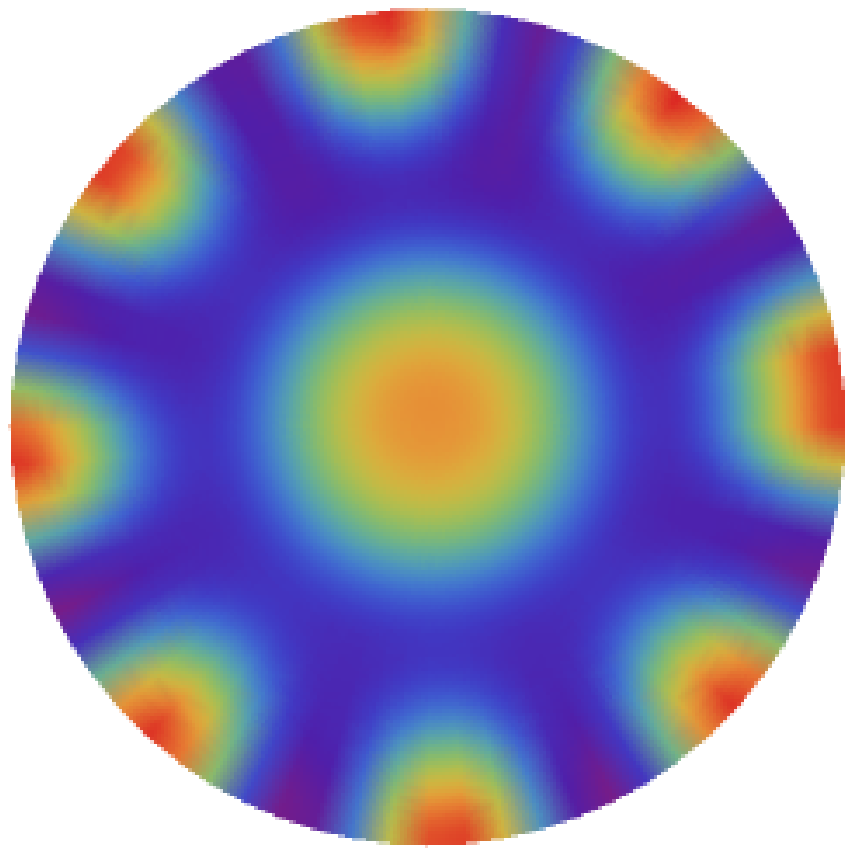}}
\caption{(Color online) Centrosymmetric  Turing patterns  under  subdiffusion conditions: $\lambda=0.5$. {\bf a)} Three-fold symmetric Turing pattern.  {\bf b)} Eight-fold symmetric Turing pattern.}
\label{fig:5} 
\end{figure}

Depending on the kinetic parameters $r_1$ and $r_2$, Turing patterns in the {\bf BVAM} model under normal diffusion conditions are either spots or stripes; the quadratic term $r_2$ favors spots while the cubic term $r_1$ produces stripes ~\cite{BarrioNum}.  Using values for these kinetic parameters such that spot patterns are favoured, Barrio {\it et al.} show that hexagonal and pentagonal patterns (centrosymmetric and noncentrosymmetric) are the most pervasive symmetries for Turing patterns on circular geometries of different size. This result can be seen in Fig.(\ref{fig:2}), where the symmetries of the obtained concentration patterns are plotted as a function of disc radius. The numerical calculations made for this figure were done with an increase in $\delta$ parameter of $0.0001$ units. The sparse regions (no symmetries reported) of Fig.(\ref{fig:2}) are regions were the  patterns symmetries were not clear, and could be solutions without rotational symmetry  or more symmetric solutions with deffects; such cases were not included in the figure.\\
In Fig.(\ref{fig:3}) we show  centrosymmetric   pentagonal and hexagonal Turing patterns, and two new symmetries that were not reported in Barrio's {\it et al.} work for normal diffusion conditions, {\it i.e.} two-fold and tetra symmetric Turing patterns.\\
In figure (\ref{fig:4})  we show the symmetries obtained under subdifussive conditions as a function of disc radius for the same values of $\delta$ used for Fig.(\ref{fig:2}). Notice that in both cases, Fig.(\ref{fig:4}a) and Fig.(\ref{fig:4}b), centrosymmetric patterns are more frequent than their normal diffusive counterpart, confirming the  linear analysis result which states that it is easier to find pure symmetries (one or few excited eigenmodes) when subdiffusion is present. Moreover, notice that the first well defined rotational symmetries appear at smaller radius than in the  normal diffusion case, and that the symmetry-radius structure of Fig.(\ref{fig:2}) change completely as subdiffusion becomes stronger, with new symmetries (eigth-fold symmetry) added for the strongest subdiffusive case ($\lambda=0.5$).
In Fig.(\ref{fig:5}) we show three and eight-fold symmetric Turing patterns obtained under the subdiffusive regime. Notice that because the eigenfunctions with $\Omega \neq 0$ are  zero at the origin, the eight-fold symmetric Turing pattern must be a mixture of more than one unstable eigen-mode. This is also the case for the five and six-fold symmetric Turing patterns of Fig.(\ref{fig:3}a) and Fig.(\ref{fig:3}b)  for normal diffusion. For a more thorough discussion about this point see ~\cite{Barrio}.
\subsection{Superdiffusion}\label{ssec:3}
\begin{figure} 
\centering
\subfloat[$\lambda=-0.66$, $\delta=0.0041$] {\includegraphics[scale=0.7]{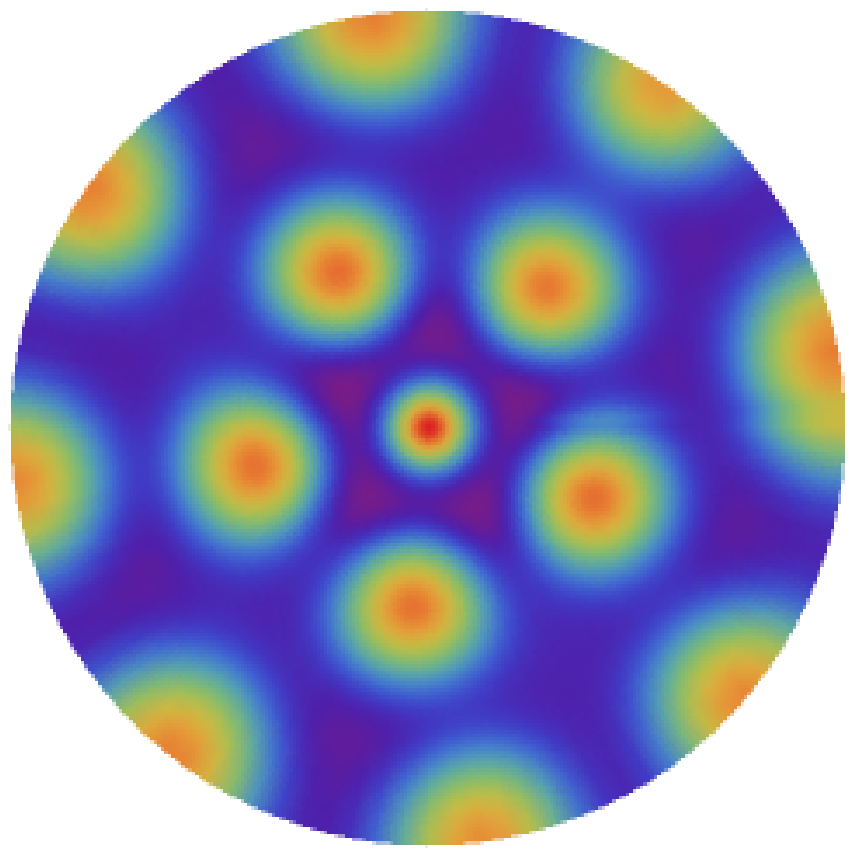}}\hspace{0.6cm}
\subfloat[$\lambda=-1\text{.}2$, $\delta=0.0007$] {\includegraphics[scale=0.7]{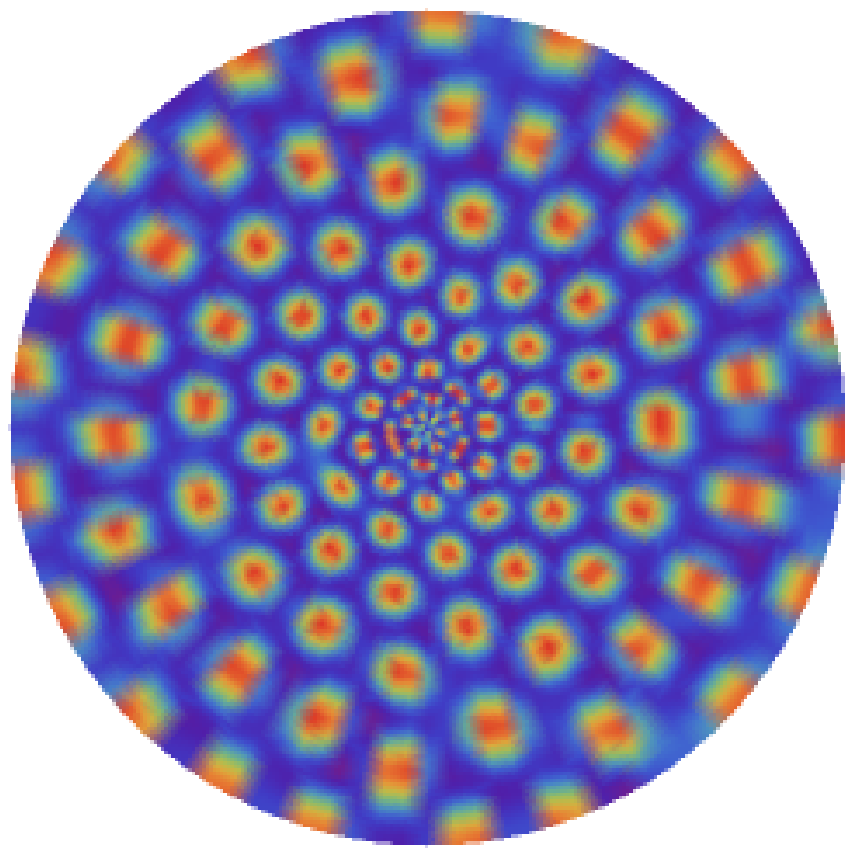}}
\caption{(Color online) Turing patterns  under  superdiffusion conditions. {\bf a)}  Turing patterns with mixed symmetries.  {\bf b)} spiral-like Turing pattern.}
\label{fig:6}
\end{figure}

\begin{figure}[htbp]
\centering
\subfloat[$\delta=0.0053$] {\includegraphics[scale=0.7]{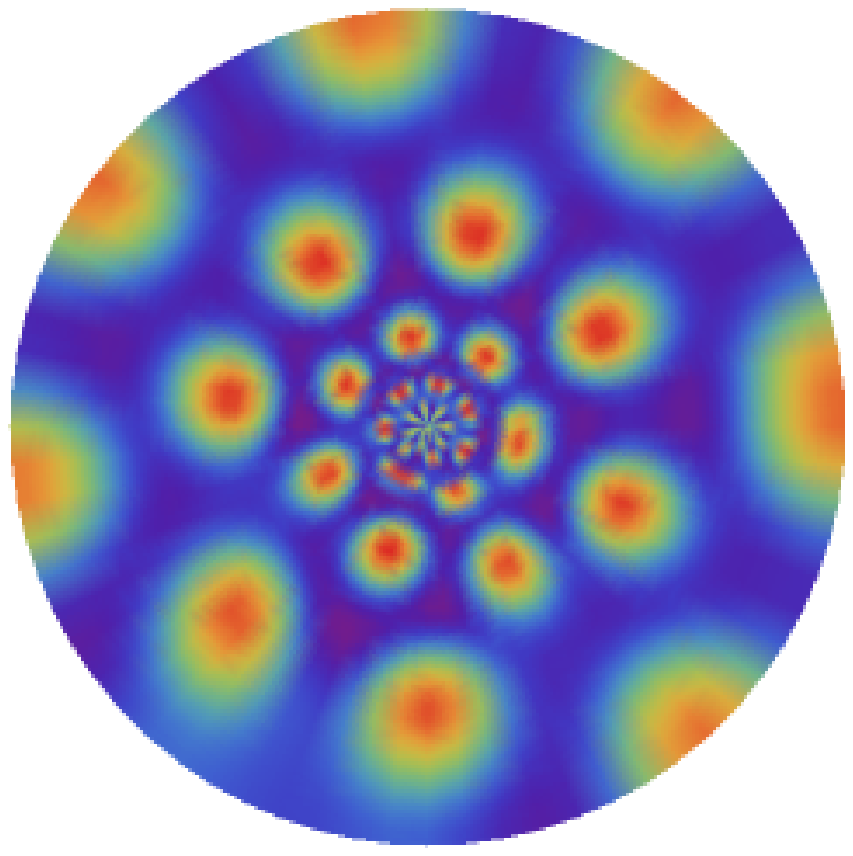}}\hspace{0.6cm}
\subfloat[$\delta=0.0074$] {\includegraphics[scale=0.7]{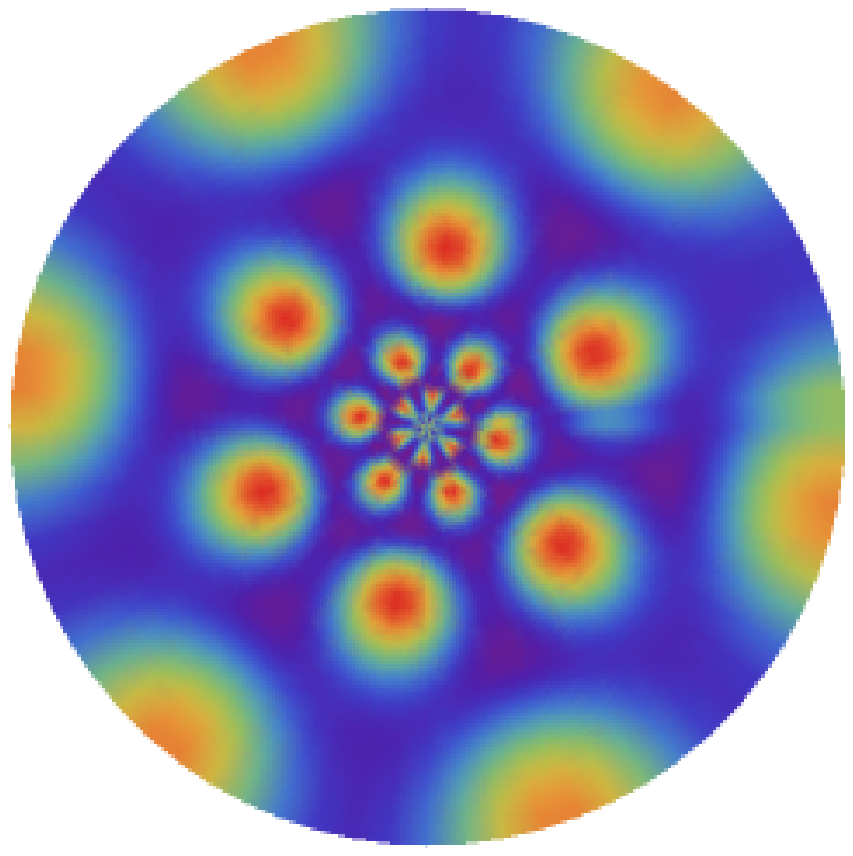}}
\caption{(Color online) Turing patterns  under strong  superdiffusion,  $\lambda=-1.95$.  {\bf a)}  Spiral like pattern.  {\bf b)} Self-similar hexagonal Turing pattern}
\label{fig:7}
\end{figure}

\begin{figure} 
\centering
\subfloat[$\delta=0.0107$] {\includegraphics[scale=0.7]{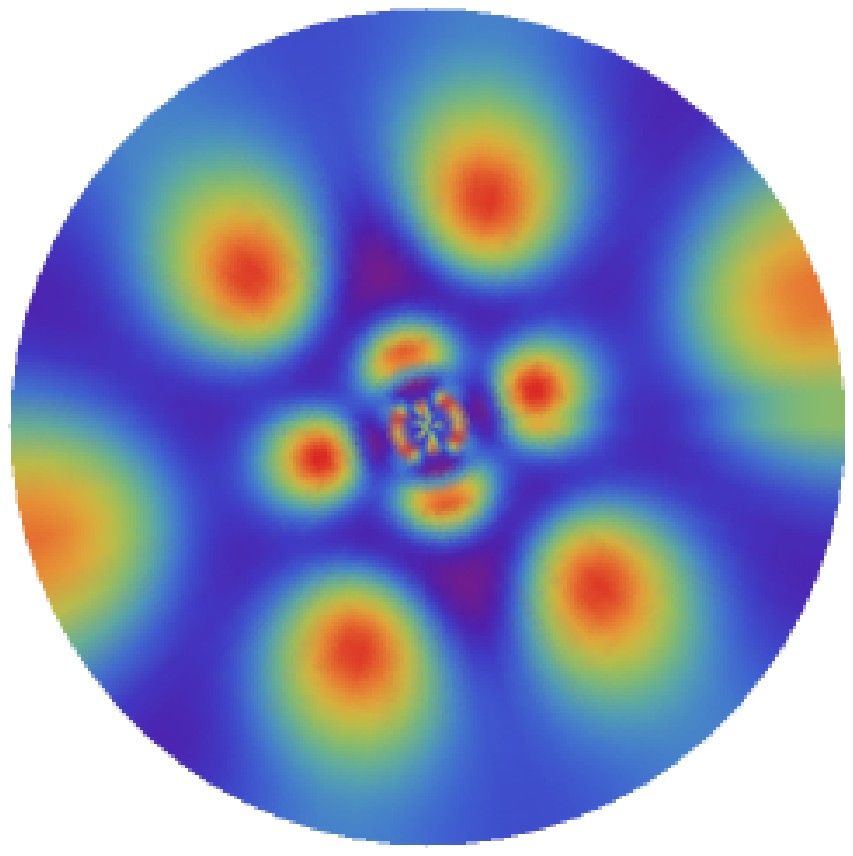}}\hspace{0.5cm}
\subfloat[$\delta=0.0110$] {\includegraphics[scale=0.7]{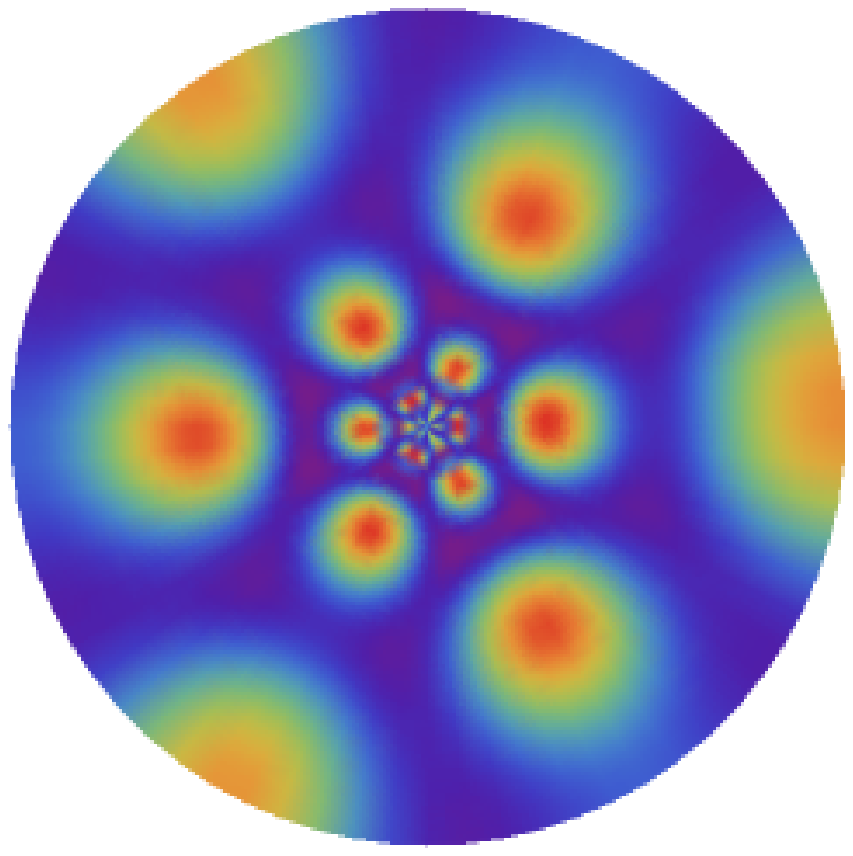}}\\
\subfloat[$\delta=0.0069$] {\includegraphics[scale=0.7]{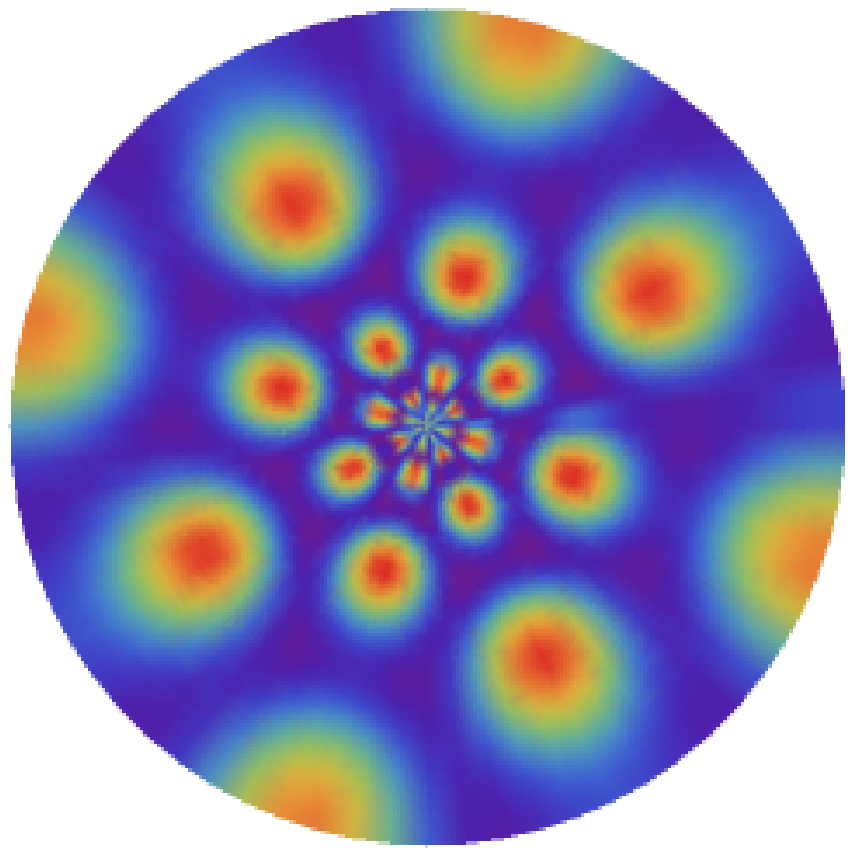}}\hspace{0.5cm}
\subfloat[$\delta=0.0115$] {\includegraphics[scale=0.7]{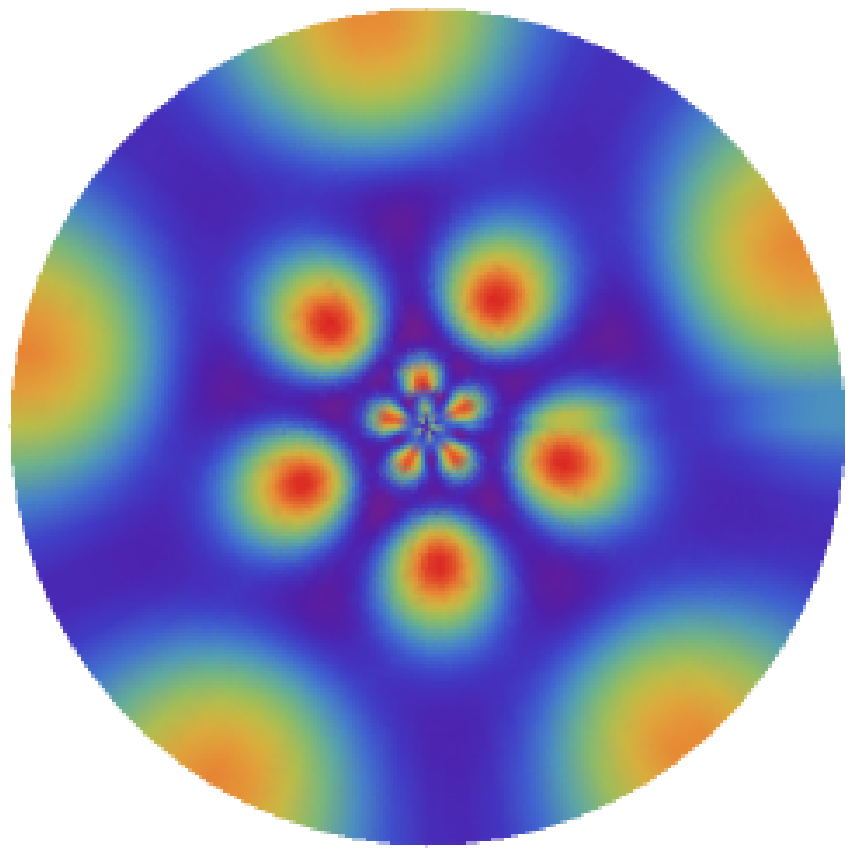}}
\caption{(Color online) Centrosymmetric self-similar Turing patterns under strong superdiffusion conditions. For each figure $\lambda=-1\text{.}95$. {\bf a)} Two-fold self-similar symmetric Turing pattern.  {\bf b)} Three-fold self-similar symmetric Turing pattern.   {\bf c)}  Tetra symmetric self-similar Turing pattern. {\bf b)} Five-fold symmetric self similar Turing pattern.}
\label{fig:8}
\end{figure}

\begin{figure}[htbp]
\centering
\subfloat[$\delta=0.0053$] {\includegraphics[scale=0.7]{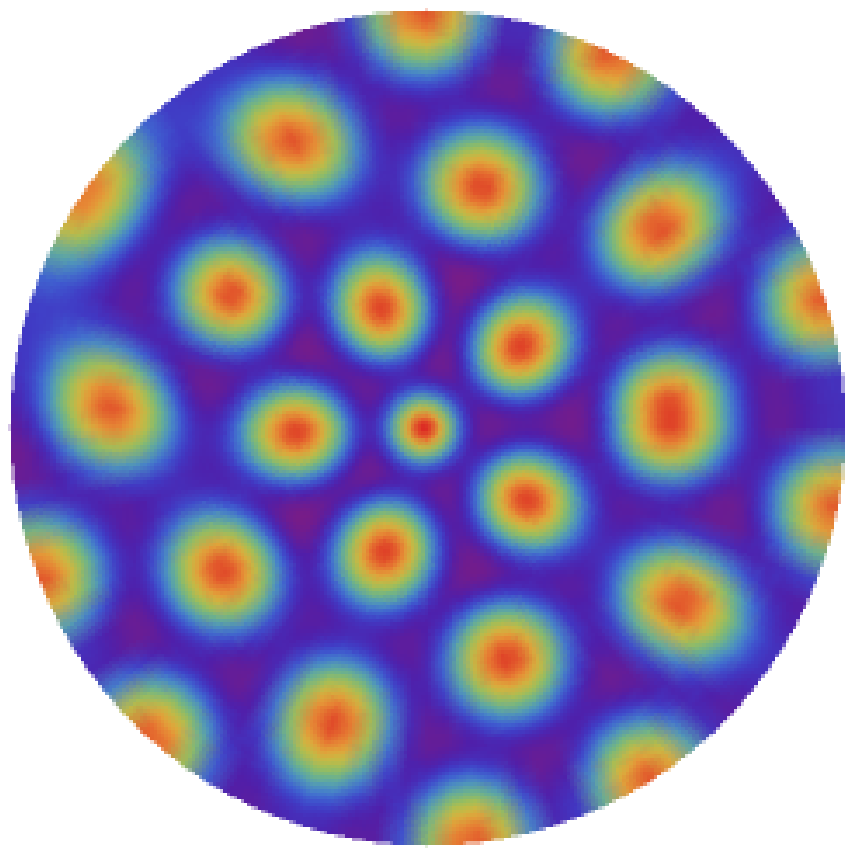}}\hspace{0.6cm}
\subfloat[$\delta=0.0074$] {\includegraphics[scale=0.7]{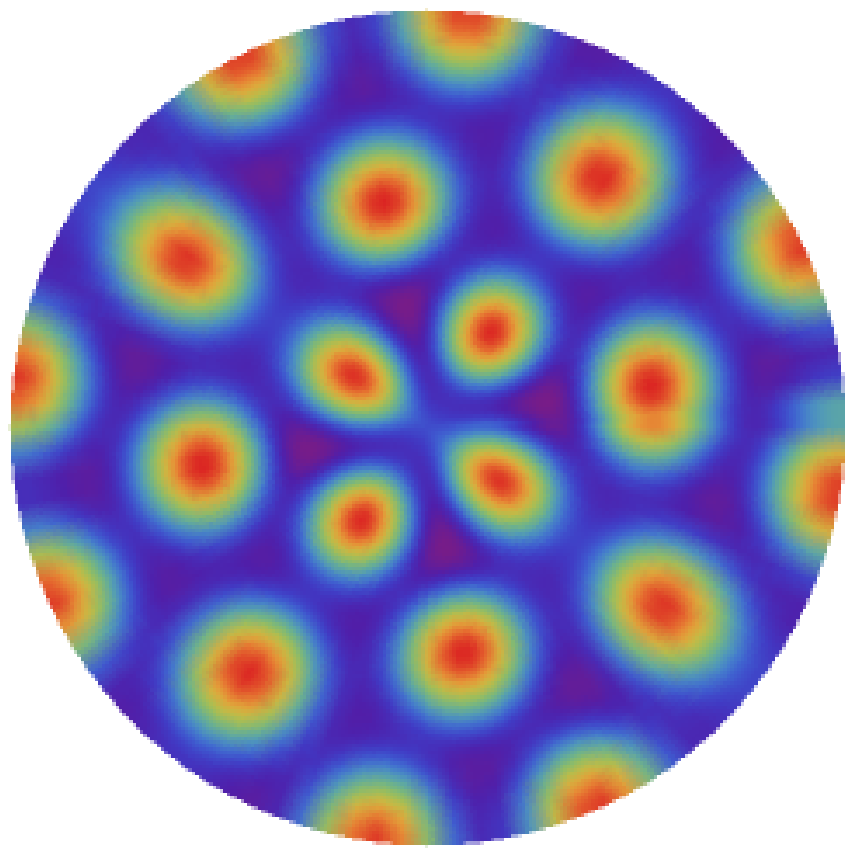}}
\caption{(Color online) Turing patterns  under  superdiffusion,  $\lambda=-0.5$.  {\bf a)} Almost five-fold centrosymmetric pattern with no repetition of the same motif for $\delta=0.0018$.  {\bf b)} Tetra-fold centrosymmetric pattern with no repetition of the same motif  for $\delta=0.0023$}
\label{fig:9}
\end{figure}

When the system is in the superdiffusive regime, {\it i.e.} when $\lambda < 0$, stationary concentration patterns with rotational symmetries are less common. However, for weak superdiffusive regimes, {\it i.e.} for $\lvert \lambda \rvert < 1 $ and for the $\delta$ parameter explored values,  many of them still conserve some of the original symmetries of the stationary steady state, {\it i.e.} reflection symmetries. On the other hand, as the superdiffusive regime becomes stronger, some interesting mixtures of symmetries and spiral-like patterns appear, see Fig.(\ref{fig:6}), and most importantly, scale-invariant or self-similar Turing patterns emerge, see Fig.(\ref{fig:7}) and (\ref{fig:8}). The fact that Turing systems under appropriate conditions can produce well organized self-similar concentration patterns is an interesting result, since to our knowledge, all  physicochemical systems capable of producing structure at a wide range of length scales, produce only structures with statistical self-similarities or disordered fractals; think for example in critical phase transitions or the percolation and diffusion limited aggregation models ~\cite{Havlin}. In the context of physicochemical systems which produce structure through spontaneous symmetry breaking, this is also the case, see for example Sharon {\it et al.} study of fractal patterns on thin sheets and biological membranes ~\cite{Sharon}.\\
Also in the superdiffusive regime, we found centrosymmetric (or almost centrosymmetric) patterns which did not repeat the same motif at different scales, examples of this kind of patterns can be seen in Fig.(\ref{fig:9}).\\
Finally, it is worth mentioning that in the superdiffuve regime, our linear stability analysis is only valid for $\lambda \in (-2,0]$. For $\lambda=-2$,  Berding and colleagues ~\cite{Berding}, showed that the eigenfunctions of the generalized diffusion equation, the equivalent of Eq.(\ref{eq:eigdif}) for $\lambda=-2$, in an annular domain with zero flux boundary conditions, are
\begin{equation}\label{eq:eigdifI}
\begin{split}
&\phi_{m,n}(r,\theta,t)\propto e^{-D_{u}k_{m,n}t}\frac{1}{r}\cos{\Bigl(nln(r)\Bigr)}\Biggl(e^{im\theta}+c.c.\Biggr),\\
&k_{m,n}=n^2+m^2+1.
\end{split}
\end{equation}
with $n$ and $m$ integers.  Interestingly, in this work, the authors also show that the level curves of Eq.(\ref{eq:eigdifI}) are logarithmic spirals.
\section{Conclusions}\label{sec:4}
Opposite to many physicochemical systems that are capable to generate structure at a wide range of length scales, structures which are commonly  considered as disordered pre-fractals, this work shows that some diffusion limited chemical reactions under specific conditions can produce well ordered, self-similar stationary  concentration patterns.
Although this exotic emergent structures are consequence of a special type of superdiffusion generated by the medium heterogeneity, which is still open to debate and experimental verification, we believe this is an important result, since in many real systems, specially in biology, self-similar well ordered structures are common.
%

\end{document}